\pdfoutput=1 

\documentclass[10pt]{article}

\usepackage{amsmath}
\usepackage{amssymb}

\usepackage{cite}

\usepackage{hyperref}
\usepackage{lineno}

\usepackage{microtype}
\DisableLigatures[f]{encoding = *, family = * }

\usepackage{amsfonts,url,float,color}
\usepackage[labelfont=bf,labelsep=none,justification=raggedright]{caption} %yet another hack to adhere to PONE requirements
\usepackage{overpic}
\makeatletter
\renewcommand{\@biblabel}[1]{\quad#1.}
\makeatother

\date{}

\newcommand{\RR}{{\color{black}R}}

\newcommand{\FF}{{\color{black}F}} 
 
\newcommand{\BB}{{\color{black}B}} 
\renewcommand{\SS}{{\color{black}\Sigma}}
\newcommand{\mm}{{\color{black}\mu}} 
\begin{document}

%CPC: line numbering requested on manuscripts http://www.plosone.org/static/guidelines.action, but template provided includes package, but does not put required inline command in.  modify here.
% \setpagewiselinenumbers %this option resets the line count on each page
% \modulolinenumbers[5]
% \linenumbers %can turn off in appendix by using \nolinenumbers
\nolinenumbers
\renewcommand{\thefootnote}{$\dagger$} 
% Title must be 150 characters or less
\begin{flushleft}
{\Large
\textbf{Inferring Latent States and Refining Force Estimates via Hierarchical Dirichlet Process Modeling in Single Particle Tracking Experiments \footnote{Differs only typographically from PLoS One article available freely as an open-access article at: \newline \url{http://journals.plos.org/plosone/article?id=10.1371/journal.pone.0137633}}}
}
% Insert Author names, affiliations and corresponding author email.
\\
Christopher P. Calderon$^{1,\ast}$, 
Kerry Bloom $^{2}$, 
\\
\bf{1} Ursa Analytics, Denver, CO, USA
\\
\bf{2} Department of Biology, University of North Carolina, Chapel Hill, NC, USA
\\
$\ast$ E-mail: Corresponding Chris.Calderon@UrsaAnalytics.com
\end{flushleft}

% Please keep the abstract between 250 and 300 words
\section*{Abstract}

Understanding the basis for intracellular motion is critical as the field moves toward a deeper understanding of the relation between Brownian forces, molecular crowding, and anisotropic (or isotropic)  energetic forcing.  Effective forces and other parameters used to summarize molecular motion change over time in live cells due to latent state changes, e.g., changes  induced by  dynamic micro-environments, photobleaching, and other heterogeneity inherent in biological processes.  This study discusses limitations in currently popular analysis methods (e.g., mean square displacement-based analyses) and how new techniques can be used to systematically analyze Single Particle Tracking (SPT) data experiencing abrupt state changes in time or space. The approach is to track GFP tagged chromatids in metaphase in live yeast cells and quantitatively probe the effective forces resulting from dynamic interactions that reflect the sum of a number of physical phenomena. State changes can be induced by various sources including: microtubule dynamics exerting force through the centromere, thermal polymer fluctuations, and DNA-based molecular machines including polymerases and protein exchange  complexes such as chaperones and chromatin remodeling complexes. Simulations aiming to show the relevance of the approach to more general SPT data analyses are also studied. Refined force estimates are obtained by adopting and modifying a nonparametric Bayesian modeling technique, the Hierarchical Dirichlet Process Switching Linear Dynamical System (HDP-SLDS), for SPT applications. The HDP-SLDS method shows promise in systematically identifying dynamical regime changes induced by unobserved state changes when the number of underlying states is unknown in advance (a common problem in SPT applications).  We expand on the relevance of the HDP-SLDS approach, review the relevant background of Hierarchical Dirichlet Processes, show how to map discrete time HDP-SLDS models to classic SPT models, and discuss limitations of the  approach. In addition, we demonstrate new computational techniques  for tuning   hyperparameters and for checking the statistical consistency of model assumptions directly against individual experimental trajectories;  the techniques circumvent  the need for ``ground-truth'' and/or subjective information.

   \newpage

\section*{Introduction}

Recent advances in optical microscopy \cite{Arhel2006, Brandenburg2007,Lessard2007, Nagerl2008, Huang08022008, Rohatgi2007,manley2008,  Pavani2009,Rohatgi2009,Thompson2010, Sahl2010, Grunwald2010,Ram2012,Gao2014,Welsher2014,Chen2014}  have inspired numerous analysis methods  aiming to quantify the motion of individual molecules in live cells \cite{Masson2009,Berglund2010,Ensign2010,Weber2012,Persson2013,Chen2013a,Calderon2013b,Stephens2013a,Aguet2013,Hajjoul2013,MassonJB2014,arxivDec2013,Chenouard2014}. The resolution afforded by current optical microscopes allows researchers to more reliably measure two-dimensional (2D) \cite{Masson2009,arxivDec2013,MassonJB2014}   and three-dimensional (3D) \cite{Calderon2013b}  position vs.~time data  in Single Particle Tracking (SPT) experiments. This permits researchers to probe \emph{in vivo} forces without introducing external perturbations into the system. Techniques capable of reliably quantifying the \emph{in vivo} forces experienced by single-molecules (without ensemble averaging) offer the 
potential to gain new molecular-level understanding of various  complex biological processes 
including  cell division   \cite{Stephens2013a}, virus assembly \cite{Jouvenet2011},  endocytosis \cite{KrapfCCP2013} and drug delivery \cite{Welsher2014}.

In this article, we demonstrate how the Hierarchical Dirichlet Process Switching Linear Dynamical
System (HDP-SLDS) framework developed by Fox and co-workers \cite{Fox2011}  can be used to deduce the direction and magnitude of different forces that contribute to molecular motion in living cells   \cite{Calderon2013b}. The utility of combining the HDP-SLDS with SPT was motivated by experiments aiming to quantify the time varying forces driving chromosome dynamics.  The approach presented shows promise in both (I) accelerating the scientific discovery process (i.e., statistically significant changes in dynamics can be reliably detected) and (II) automating preprocessing tasks required when analyzing and segmenting large SPT data sets.  

The technique introduced is applicable to various scenarios where SPT trajectories are sampled frequently in time and particles can be accurately tracked over multiple frames, e.g. \cite{arxivDec2013,Calderon2013b,Welsher2014,Hajjoul2013,Chen2014}.  
Extracting accurate and reliable force estimates from noisy position vs.~time  data in the aforementioned setting requires one to account for numerous complications inherent to experimental SPT data in living cells. For example, nonlinear and/or time changing systematic forces need to be differentiated from thermal fluctuations (i.e., random forces), both of which contribute to motion at the length and time scales measurable in living systems \cite{Calderon2013b,KrapfCCP2013,Stephens2013a}. Furthermore, additional measurement noise (consisting of localization error amongst other factors \cite{Thompson2002,Enderlein2006,Gahlmann2013,Berglund2010,Calderon2013b}) induced by the optical measurement apparatus  must be systematically accounted for since this noise source varies substantially between and within single trajectories; inaccurate effective measurement noise estimates can appreciably influence estimates of kinetic parameters   as well as statistical decisions about the underlying physical system \cite{Berglund2010,Calderon2013,Calderon2013b,arxivDec2013}.
Finally, extracting forces from position vs.~time data requires one to explicitly or implicitly make numerous assumptions about the underlying effective dynamics.  We believe these assumptions should be systematically tested directly against experimental data before one trusts kinetic quantities inferred from experimental data \cite{SPAfilter,SPAdsDNA,Calderon2013b}.  However, in live cell SPT studies, reference ``ground-truth'' is rarely available. Hence, techniques  checking statistical assumptions directly against data are attractive (e.g., through goodness-of-fit hypothesis testing \cite{Calderon2013b}).  

The HDP-SLDS approach  combines Hidden Markov Modeling (HMM),
Kalman filtering, \cite{hamilton,stengel1994}, and more recent ideas from Dirichlet Process modeling \cite{Teh2006a}. 
We demonstrate how the HDP-SLDS 
method 
can be used to reliably identify the time at which a state change occurs as well as the number of  states implied by a specific time series.   In the HDP-SLDS approach \cite{Fox2011}, the number of underlying states are inferred 
from the data via nonparametric Bayesian techniques  \cite{GhoshNonparBook,Teh2006a}.  Inferring the number of states jointly  with the parameters determining the dynamics (i.e., in a single fully Bayesian computation) is useful because the number of underlying effective states is rarely known \emph{a priori} in live cell SPT applications due to inherent heterogeneity between and within trajectories \cite{Persson2013,Calderon2013b,Chen2013a}.  Since the HDP-SLDS framework directly infers the number of states from observed data, the user does not need to provide an accurate upper bound on the number of states or worry about \emph{a posteriori} model selection issues \cite{claeskens_modselec_BOOK,Persson2013,Schwantes2014}. This  allows the HDP-SLDS approach to readily identify a wide range of distinct dynamical regimes which may occur within a single trajectory. 
Once the experimental trajectory is segmented into distinct kinetic states, one can use classic maximum likelihood estimation (MLE) techniques to infer kinetic parameters \cite{llglassy,Calderon2013b}. Carrying out MLE estimation after segmentation mitigates the bias induced by prior assumptions.
It is emphasized throughout that the procedure described in this work can systematically account for measurement noise and complex spatial-temporal  variations in thermal fluctuations \&  force/velocity fields.

In the particular application that we studied, the heuristics suggested for generating HDP-SLDS prior parameter input required modification before accurate results could be obtained in SPT applications.  Specifically, prior specification  heuristics employed  by the original HDP-SLDS Bayesian analysis (using guidelines outlined in Ref. \cite{Fox2011}) required substantial modification 
(prior specification substantially influences both state inference and molecular motion parameter estimation).
In particular, the ``hyperparameters" governing the measurement and thermal noise (parameterizing components of the
``base-measure'' \cite{Fox2010,Fox2011}) required more careful calibration/tuning before high accuracy results could be obtained.  In S1 Text,
we discuss how  ideas in Refs. \cite{Berglund2010,Calderon2013b}  can be used for this type of hyperparameter tuning.

We note that the experimental applications are meant to serve as a proof-of-concept. The results presented are not intended to provide an exhaustive study of the forces involved in chromosome movement in metaphase.  The problem is amenable to the HDP-SLDS approach since the GFP tagged experiments experienced  abrupt state changes   in the experimental trajectories before other downstream SPT computations could be reliably carried out (e.g., force and diffusion coefficient estimation).  Not all ``abrupt'' were visually obvious to human observers, hence the approach is promising in both  data-mining and preprocessing contexts.  
It should also be noted that two recent articles have discussed the utility of nonparametric Bayesian ideas in single-molecule data analysis 
\cite{Calderon2014,Hines2015}; Ref. \cite{Calderon2014} is similar in spirit to this manuscript except the focus is exclusively on simulation data mimicking features of SPT data and Ref. \cite{Hines2015} focuses on using nonparametric Bayesian modeling to detect the number of states in HMM models.  Note that in Ref. \cite{Hines2015} each observation is assumed to be drawn independently from a distribution indexed by a latent state and the approach employed does not assume explicit temporal dependence between successive SPT time series observations  whereas this work accounts for such temporal dependence  
(allowing quantitative approximations of spatially or temporally dependent effective forces and/or  velocities).

     This article is organized as follows:  the mathematical models and experimental methods used are presented in Sec.~\ref{sec:meth};
       Results obtained when analyzing chromatid dynamics in yeast  are presented in Sec.~\ref{sec:expres}.  Conclusions are presented in Sec.~\ref{sec:discussion}. 
     We have also included Supporting Material (S1 and S5 Texts) where additional technical details are provided and control simulation studies are analyzed (results shown in S1-S7 Figs.).  A companion publication  \cite{Calderon2014} compares the HDP-SLDS approach to an open-source SPT Hidden Markov Model assuming a finite number of states through simulation studies.  

\section{Materials and Methods}
\label{sec:meth}
   
\subsection{Background and Models Considered}
     We begin by reviewing the main technical ideas underlying the HDP-SLDS 
    introduced in Ref. \cite{Fox2011}.  Note that the SLDS  models assumed allow researchers to decouple noise induced by
    thermal fluctuations, measurement noise, and spatially anisotropic velocity/forces. Forces are computed by leveraging the  overdamped Langevin model \cite{Calderon2013b}  (this relationship is made explicit in Sec. \ref{sec:ODdesc} and S3 Text).  

    In the early SPT works, the spatial and temporal resolution afforded by the measurement device led researchers to focus mainly on Mean-Square-Displacement (MSD) type analyses to analyze single-molecule data
\cite{Kusumi1993,Saxton1997a,Golding2004,Park2010}.  MSD approaches have many undesirable features, namely they tend to introduce unnecessary temporal averaging (i.e., they ignore the natural time ordering of the trajectory measurements) and they have a difficult time accounting for spatially varying forces (a common occurrence in live cells \cite{Calderon2013b}).   Advances in spatial and temporal resolution have inspired many researchers to develop new techniques aiming at more reliably extracting single-molecule level information out of measurements \cite{Montiel2006,SPAfilter,SPAdsDNA,Ensign2010,Presse2013,Persson2013,Chen2013a,Calderon2013,MassonJB2014,VandeMeent2014,Schwantes2014}.  The previously cited works are most similar in spirit to the work presented, but all of the works encounter technical difficulties when the number of ``states'' is not known in advance. 
 Additional technical complications arise when position estimates are obscured by non-negligible ``measurement noise" \cite{SPAfilter,Berglund2010} or force/velocity fields exhibiting temporal dependence \cite{Calderon2013b}.

      The method of Fox et al. \cite{Fox2011} overcomes the difficulties mentioned above by assuming that a discrete time series model of the form:

    \begin{align}
    \label{eq:SDE}
    \vec{r}_{i+1}= & \vec{\mm} + \FF \vec{r}_i+\vec{\eta}_i; \ \ \vec{\eta}_i \sim \mathcal{N}(0,\SS) \\
    \label{eq:SDEm}
    \vec{\psi}_{i+1}= & \vec{r}_{i+1} +\vec{\epsilon}_{i+1};  \ \ \vec{\epsilon_{i+1}} \sim \mathcal{N}(0,\RR), \  
    \end{align}

    \noindent can be used to describe the  dynamics of each unique state. The position of the molecule or particle at time $t_i$ is denoted by the vector $\vec{r}_{i}$ and the measured value of 
    the position at this same time is denoted by $\vec{\psi_{i}}$ (subscripts are used to index time);  the position is not directly measurable due to  ``localization noise'' and
    other artifacts induced by the experimental apparatus. The term ``effective measurement noise'' is meant to include the net measurement
    noise induced by finite photon counts, background fluorescence, motion blur, etc.
    \cite{Berglund2010,SPAfilter,Calderon2013b, Gahlmann2013}.  Effective measurement noise is modeled as a mean zero normal random variable with covariance $\RR$; the expression
    $\vec{\epsilon} \sim \mathcal{N}(0,\RR)$ conveys that the random vector, $\vec{\epsilon}$, is distributed according to 
    the normal distribution $\mathcal{N}(0,\RR)$.  Techniques for checking the validity of this modeling assumption via hypothesis testing technique are discussed in Ref. \cite{Calderon2013b}. 

    The term $\vec{\mm}$ represents a  (constant)  ``velocity vector" experienced by the particle multiplied by the observation time $\Delta t$;  the
    matrix $\FF$ accounts for systematic spatial variations in forces or velocity; random thermal fluctuations are 
    modeled by $\vec{\eta}$.  The ``drift terms'' \cite{SPA1,SPAfric}, i.e. $\vec{\mm}$ and $\FF \vec{r}_i$, can be used to quantify active (linear) force and velocity fields; %(see \ref{sec:techdetails1}); 
    spatial anisotropy in the force and velocity fields are accounted for by $\FF \vec{r}_i$. The physical assumptions we employ for inferring forces from position vs.~time data are discussed in Sec. \ref{sec:ODdesc}. 
    %(for precise mathematical relations required to back out forces from the HDP-SLDS model parameters,  see \ref{sec:techdetails1}).
    The net parameter vector characterizing the (discrete) dynamics of a single state is given by $\theta=(\vec{\mm},\FF,\RR, \SS)$. Note that the HDP-SLDS presented in Ref. \cite{Fox2011} assumed that all observations are uniformly spaced by $\Delta t$ time units.

    Although the molecular position, $\vec{r}$, is not directly observable,  
    the discrete state-space model above allows one to readily plug directly into the established Kalman filtering equations 
    \cite{hamilton,stengel1994} to infer the dynamics of $\vec{r}$ using the observations $\vec{\psi}$.  The Kalman filtering framework allows one to systematically treat measurement noise as well as
    spatially dependent particle forces and/or velocities  (this spatial variation can induce classic ``confinement'' or ``corralling'' effects or be indicative of molecular binding in SPT data \cite{Calderon2013,Calderon2013b}). The importance of properly accounting for measurement noise is demonstrated  in S2 Text and is also presented 
    elsewhere \cite{SPAdsDNA,Berglund2010,Calderon2013,Calderon2013b}.  

     An  illustrative trajectory of  the (observable) measurements $\vec{\psi}$ and (unobservable) $\vec{r}$ is shown in Fig. \ref{fig:illEG};  this figure will also be used to also illustrate the ``latent state'' modeling discussed in the next paragraph.
     Note that the presence of
    $\FF$ slightly complicates physical interpretation of the parameters \cite{Calderon2013b}, but the discrete model above can be readily mapped to a continuous time stochastic
    differential equation (SDE) studied in Ref. \cite{Calderon2013b} (where diffusion coefficients, effective friction, and instantaneous 
    force terms associated with an overdamped Langevin equation can be readily extracted).   
    The equations mapping discrete time parameters used in the HDP-SLDS framework to continuous time SDE parameters are presented  in  S3 Text for the reader's convenience.  The ability to map between discrete and continuous time models is important for both specifying physically inspired priors and for interpreting parameter estimates since classical quantities like diffusion coefficients and forces are typically defined via continuous time SDE models \cite{Masson2009,Berglund2010,Calderon2013b}. 

    In the discrete time model above,  $\theta$ contains the parameters (vectors and matrices) required to specify  the stochastic dynamics defining the evolution equations of the system in a single ``state''.  However, simple linear evolution equations are not expected to be valid for the entire duration of the trajectory in live cell measurements \cite{Calderon2013b}.
    In this article, we assume that $\theta$  can change abruptly over time or space and hence the system's ``state" 
    can change over time.  
    We
    will use the notation $z_i$ to denote the state  at time $i$ and will use $\theta_{z}$ to denote the parameter 
    vector characterizing the 
    dynamics of $\vec{r}$ when it is in
    state $z$.  The standard goal of an HMM inference procedure is to infer the state sequence $\{z_1,z_2, \ldots z_T \}$ from an observation sequence 
    $\{\vec{\psi}_{1},\vec{\psi}_{2},\ldots \vec{\psi}_{T}\}$.

    The typical HMM framework assumes temporal transitions between the  states are governed by Markovian 
    transition probabilities.  If there are $K$ fixed states, the transition vector associated with state $i$ in the traditional HMM framework 
    is prescribed by the vector $\vec{\pi}^{(i)}=\big(\pi_1^{(i)},\pi_2^{(i)},\ldots \pi_K^{(i)} \big)$.  Note that each component provides
    the probability of the state transition from state $i$ to state $j$ and the sum of components
    of this $K$ dimensional vector is one (hence defines a proper discrete probability distribution); the collection  
    $\{ \vec{\pi}^{(1)}, \vec{\pi}^{(2)}, \ldots \vec{\pi}^{(K)}  \}$ defines the classical HMM ``transition matrix''.  
    In Bayesian inference of standard HMMs, the Dirichlet distribution is sometimes used as a prior for $\vec{\pi}^{(i)}$ \cite{Persson2013,VandeMeent2014}.  Additional details on the Dirichlet distribution and various infinite dimensional extensions, namely the Dirichlet Process and the Hierarchical Dirichlet Process, are provided in S4 Text.

    The HDP-SLDS framework of Fox et al. comprehensively addresses  issues not accounted for in Refs. \cite{Jaqaman2006,Persson2013,Chen2013a,Calderon2013b,MassonJB2014}.  The following modified quote from Ref. \cite{Fox2011} captures the essence of the HDP-SLDS method:   
\emph{``The HDP-SLDS is an [infinite discrete state space] extension of hidden Markov models (HMMs) in which each HMM state, or mode, is associated with a linear dynamical process."}  
    As in Ref. \cite{Calderon2013b},   the statistical influence of thermal fluctuations and effective measurement noise on  the dynamics is accounted for 
     using the Kalman filter framework (this allows linear spatial variations in force and/or velocity); however,
    the HDP-SLDS also provides a mechanism for segmenting trajectories into chunks where dynamics are distinct.  
    In addition, the HDP-SLDS framework advocated in Ref. \cite{Fox2011}  introduces a ``sticky parameter'' encouraging temporal state persistence;  this feature can be advantageous in many SPT applications  \cite{Calderon2014}.  The ability to avoid model selection nuances \cite{claeskens_modselec_BOOK,Persson2013,Schwantes2014} and infer the number of states in a data-driven fashion within a single Bayesian inference while also rigorously accounting for spatially or temporally varying stochastic dynamics are major advantages of the HDP-SLDS method.

     Before discussing the major technical weaknesses of the HDP-SLDS method, we discuss some strengths/weaknesses of published  SPT modeling approaches.  Wavelet methods have the ability to identify sharp and abrupt changes in time without making too many assumptions about the underlying stochastic process \cite{Chen2013a}, but they  do not readily account for subtle spatial variations in the dynamics or noise (e.g., temporally varying velocity fields or  diffusion coefficients) and can have difficulty in separating diffusive noise from measurement noise.   The HDP-SLDS overcomes the aforementioned challenges by assuming a specific parametric model 
     (however, if anomalous diffusive noise is deemed important to quantify system dynamics \cite{Hajjoul2013,Chen2013a}, this can be a problem for the HDP-SLDS approach).  
     Ref. \cite{Jaqaman2006} is one of the pioneering efforts attempting to account for spatial and temporal variations in SPT signals, but the approach neglected to explicitly account for the effects of measurement noise (the approach also focused on comparative hypothesis tests and model selection as opposed to goodness-of-fit tests directly checking the consistency of a model's distributional assumptions  to experimental data).
    The method in Ref. \cite{MassonJB2014} attempts to account for spatial variations (allowing for correlated 2D forces) and the influence of measurement noise, but appeals to \emph{ad hoc} statistical approximations which can adversely affect state and parameter inference \cite{Berglund2010};  these approximations (which can be avoided using Kalman filtering ideas \cite{Calderon2013b}) can substantially complicate downstream analysis where one would like to check the assumptions against experimental data \cite{Calderon2013} (also Ref. \cite{MassonJB2014} did not employ any formal hypothesis testing procedures).  
    The ``Windowed local MLE" method in Refs. \cite{llglassy,SPAfilter} can account for spatial variations in force and measurement noise \cite{Calderon2013b,arxivDec2013} and provides a procedures for goodness-of-fit testing, but does not prescribe a systematic and automated method for segmenting time series data into distinct dynamical segments.

    Open problems facing the HDP-SLDS are associated with  prior specification and model validation \cite{Fox2010}. The priors used are selected primarily for computational convenience.  For example, the so-called matrix normal inverse-Wishart (MNIW) and other priors proposed allow for exact Markov chain Monte Carlo (MCMC) sampling \cite{Fox2011}, but such priors have no real connection to the physical mechanisms governing molecular motion.  The priors and  their associated hyperparameters used by the HDP-SLDS method are discussed further in S5 Text.  In the results, we demonstrate how a variant of the method in Ref.    \cite{Calderon2013b} can be used to correct for some artifacts 
     introduced by ``bad   priors'' (i.e., the parametric prior distributions assumed do not accurately reflect the true underlying process).
     Quantitatively prescribing a ``good prior'' is difficult because single-molecules trajectories contain a high degree of heterogeneity 
     induced by the local micro-environment, conformational fluctuations of the biomolecules, etc.  Hence finding an accurate prior representative of a single trajectory is non-trivial  (finding a prior governing a population of trajectories in the spirit of Ref. \cite{VandeMeent2014} is  even harder in  SLDS modeling).  
     Fortunately, the nonparametric Bayesian HDP-SLDS from Ref. \cite{Fox2011} combined with frequentist ideas   \cite{Calderon2013b}  provides researchers in SPT analysis a set of tools which can be used to more reliably and  systematically extract information from complex \emph{in vivo} measurements as we demonstrate in the Results and Supporting Text. 

     \subsubsection{Inferring Instantaneous Forces from Noisy Position Measurements}
     \label{sec:ODdesc}

In what follows, 
we present a technical discussion explicitly pointing out how we infer forces from a sequence of position measurements.  Particle position is modeled using the overdamped limit of the Langevin equation (i.e., a first order SDE). The general form of the corresponding first order SDE is:
\begin{align}
\label{eq:ODg}
d\vec{r}_t = \gamma ^{-1} \vec{f}(\vec{r}_t)dt + \sqrt{2D}d\vec{B}_t,
\end{align}
\noindent where $\gamma$ is a local effective ``friction matrix" associated with the Langevin model and $\vec{f}(\cdot)$ represents the force vector \cite{Calderon2013b,MassonJB2014}; within the SLDS the force is a function of both the position $\vec{r}$ and the latent state.  The notation above was selected to facilitate comparison to Refs. \cite{MassonJB2014} and \cite{Hoze2014}. 
The above model assumes particle inertia can be ignored 
(i.e., the dynamics are in the ``overdamped" regime or the Smoluchowski limit of the full Langevin equation \cite{Hoze2014}) and local diffusivity is constant.  In Text S3,  the mapping between the continuous and discrete HDP-SLDS time series models is provided (note that in S3 Text,  
$\gamma^{-1} \equiv \Phi$). 
On the time and length scale of SPT measurements, it is generally accepted that particle inertia is unimportant.  However, goodness-of-fit tests can be used to check if unobserved and unmodeled momentum dynamics affect position measurements \cite{SPAgof,arxivDec2013}.  If inertia is believed or empirically determined to be important, more complex (second order SDE) models for fitting forces from position data can be entertained \cite{stuart_hypoelliptic_09}.

The subtler assumption implicit in our estimate of the forces from the above model is that the Einstein %(or fluctuation dissipation)
 relationship can be used to relate local effective friction to the local effective diffusion coefficient via the relationship:
\begin{align}
\label{eq:FD}
\gamma ^{-1} = \frac{D}{k_BT}.
\end{align}
The above relationship is a classic microscopic relationship relating molecular diffusivity to local friction \cite{gardiner,Park2004,Hoze2014}. On longer time scales, the microscopic parameters become effective parameters \cite{Calderon2013b,Hoze2014} and the validity of the above relationship can become questionable due to phenomena like molecular crowding 
\cite{Holcman2011} or other factors (e.g., transient binding) inducing more complex dynamics.
The temporal segmentation approach proposed aims to detect when a sudden event alters the dynamics in a statistically significant fashion, hence the odds of ``mixing" different dynamical regimes is reduced (e.g., if the obstacle density experienced in one part of the cell is vastly different than another, then the effective dynamics in each regime will be different \cite{Holcman2011}).  The HDP-SLDS's ability to identify regimes, within a single trajectory, where locally linear SDE models are consistent with the data
 makes the validity of Eq. \ref{eq:FD} more plausible.   

It should be noted that ``effective force" data has been backed out from noisy position measurements in multiple \emph{in vitro} single-molecule manipulation experiments \cite{SPAfilter,SPAdsDNA,SPAfric}.  In the aforementioned works, force measurements obtained via an external probe (e.g., AFM) could be compared to internal forces inferred from noisy position vs.~time data (more  precisely molecular extension vs.~time) obtained by appealing to relationships similar to those in  Eqns. \ref{eq:ODg} and \ref{eq:FD}.  Despite the system's being far out of equilibrium in Refs. \cite{SPAfilter,SPAdsDNA,SPAfric},  appealing to the relationship in Eq. \ref{eq:FD} yielded effective ``internal forces'' consistent with those measured by external probes.
 However,
testing the validity of the above expression directly against \emph{in vivo} SPT data is non-trivial since directly measuring fores in the crowded and heterogeneous environment of live cells is difficult.   
The formal statistical hypothesis testing procedure employed can only  judge the consistency of the assumed drift and diffusion terms of the SDE in Eq. \ref{eq:ODg} against observational data; said differently, only the product $\gamma ^{-1} \vec{f}(\vec{r}_t)$ affects the test (the validity of Eq. \ref{eq:FD} does not influence the hypothesis test in any way).
  As we will discuss later, our force estimates obtained (without assuming or inputting a local effective cellular viscosity) are consistent with other \emph{in vivo} force measurements \cite{Fisher2009}.  Even if a researcher does not believe Eq. \ref{eq:FD} should hold, our fitted models can still produce local effective velocity fields \cite{Hoze2014} which can aid in our quantitative understanding of intracellular dynamics.

Before proceeding, we explicitly mention a few more technical notes about our nominal force estimates.  Our models make no assumptions about the existence of a gradient potential field \cite{MassonJB2014} or the stationarity (temporal or spatial) of statistics characterizing the random process \cite{Hoze2014}.  Transient structures and phenomena,  which are often resolvable in live cell SPT measurements \cite{Calderon2013b}, can invalidate the validity of a force obtained by taking the gradient of an energy potential field and can also complicate nonparametric estimators  \cite{Hoze2014} (nonparametric estimators often make implicit stationarity assumptions).  Our models sample data on millisecond timescales, but allow for the local effective forces and diffusion to change over time and/or space (i.e., statistics are allowed to be non-stationary);  we believe this feature will be critical in accurately characterizing heterogeneous high resolution optical microscopy data.

    \subsection{Experimental Methods}
    \label{sec:expmeth}

    We examined the \emph{in vivo} dynamics of chromatin during mitosis to determine the behavior of a region of the chromosome visualized through the binding of lac repressor fused to GFP (LacI-GFP) to lac operator (lacO). The lacO is a repeated array of operator DNA sequences (256 repeats, 10 kilobase prs.) integrated 6.8 kb from the centromere on chromosome XV. The spindle pole bodies (sites of microtubulare nucleation) are visualized through a fusion protein between a spindle pole component (Spc29) and RFP (red fluorescent protein). 
    Cells were grown to logarithmic phase at 24 $^\circ$C in rich media.  Images were acquired on a Nikon Eclipse Ti wide-field inverted microscope with a 100x Apo TIRF 1.49 NA objective (Nikon, Melville, New York, USA) and Andor Clara CCD camera (Andor, South Windsor, Connecticut, USA) using continuous laser illumination. 
    Images were streamed at the net effective camera acquisition rate of 22 frames/sec. The CCD camera's exposure time was set nominally to be the inverse of the net frame rate. 
    Images were acquired at room temperature with Nikon NIS Elements imaging software (Nikon, Melville, New York, USA).  The program 
    ``Speckle Tracker'' and other methods outlined in Ref. \cite{Verdaasdonk2013} were used to estimate the centroid of GFP and RFP spots for position measurements. 

% Results and Discussion can be combined.
\section{Results}
\label{sec:expres}

     In  Fig. \ref{fig:Fhist} we  display a histogram of the estimated instantaneous force magnitudes as well as the force magnitude trajectory computed from a time series of measured $X/Y$ chromatid positions (the $X/Y$ data and state estimates are shown in Fig. \ref{fig:noiseJump}).  The force was estimated by first determining the number of states implied by the observed trajectory and using the  method described in S1 Text to segment the trajectory.  For the experimental  trajectories shown in this paper,  the initial prior mean of the measurement noise covariance matrix, $\RR$, was assumed to be the identity matrix multiplied by a scalar, $\sigma^2$, where  $\sigma= 40 nm$.  This value was inspired by the fact that the effective measurement noise standard deviation was found to be in the $10-60nm$  range for these SPT trajectories (with a mode at $40 nm$).     
    We specified $K=10$ for the so-called ``weak limit approximation" of the HDP-SLDS \cite{Fox2011a}; this term is discussed further in S5 Text and the relative insensitivity of the HDP-SLDS method to the $K$ parameter is demonstrated in  Fig S5.
    Additional parameters required to run the HDP-SLDS segmentation are reported in S5 Text. 
    After the HDP-SLDS segmentation was obtained,  
    the estimated MLE parameter vector of each unique state was used to obtain  molecular position estimates via the Kalman filter \cite{hamilton,Fox2011}. 
    Finally, the position estimates along with the MLE parameters were plugged into the equations  shown in S3 Text to evaluate the instantaneous effective force (note: this provides a collection of 2D force vectors). 

    The force vector magnitudes reported in Fig. \ref{fig:Fhist}  observed are representative of other chromatid data sets studied using this analysis.  However, some datasets exhibited more interesting ``force regime changes'' (as shown and discussed in the final experimental SPT trajectory studied).   
Two things should be  emphasized in the analysis of this trajectory: (I) previously published works have reported that chromatids in  metaphase-like conditions experience forces in 
    the (relatively low) 0.1-0.2 pN range \cite{Bouck2007,Fisher2009}.  Our findings are consistent with these previous results, except the model considered
    here
    does not require an estimate of the effective viscosity (a quantity difficult to estimate in live cells) to infer force vectors.  
    Local effective forces are estimated using time series analysis techniques applied to the so-called 
    overdamped Langevin equation outlined in Sec. \ref{sec:ODdesc}.  
    (II)  The HDP-SLDS method aided in accurately determining a transition between two states (physical interpretation of the identified states discussed below). 

    The top left panel of Fig. \ref{fig:noiseJump} displays the white light image of a trajectory obtained from  the yeast chromatid SPT experiments.   Note that we use the phrase ``white light image"   throughout to indicate a single diffraction limited image of the yeast cell obtained with  white light illumination; the trajectory obtained using a laser frequency tuned to enhance GFP excitation was then overlaid upon this single image (all images are recorded at different times, but the yeast cell is not expected to move substantially during the experiment, so the white light illumination image gives one an idea of the spatial environment explored by the molecule). 
    The top right panel displays the trajectory in $X/Y$ space with time information color coded.  
    The red vertical line in the bottom left panel of Fig. \ref{fig:noiseJump} shows the time at which a state change point was detected by the HDP-SLDS method.     
    The bottom right panel of Fig. \ref{fig:noiseJump} shows that state estimates of the vbSPT  \cite{Persson2013} and HDP-SLDS methods both identified two states, though the latter captures the state persistence more accurately.

    After the temporal segmentation was formally estimated using the HDP-SLDS, we computed  the posterior mode of $\theta$ implied by the assumed HDP-SLDS model for each of the states in the two identified  temporal segments.  These two parameter vectors (one vector drawn for each of the two states) were used along with the  experimental data
    to compute the $Q$ test statistic \cite{hong} and the corresponding $p-$values.  The null hypothesis of the one-sided goodness-of-fit test using the $Q$ statistic assumes that the time series was generated by a Markovian SDE (evaluated at an ``optimal'' parameter estimate \cite{hong}) and the alternative hypothesis assumes that the time series was produced by any other process  
    \cite{hong,SPAgof,Calderon2013b}.  ``Optimal'' parameter estimates  were obtained by two means: (i) drawing the parameter vector obtained after $10^{4}$ iterations of the MCMC sampler in the HDP-SLDS inference; note that this approach included prior bias and (ii) via MLE estimation applied to HDP-SLDS segmented trajectories;  with this approach, bias is mitigated since priors do not influence the parameter estimates.

    In the first segment,  using the observed data and the HDP-SLDS parameter vector estimate mentioned above, resulted in a    
    $p-$value of 0.32 (little evidence for rejection), but in the second segment a $p-$value $<$ 0.01 was computed (large evidence for rejection). Visual inspection of Data S1 (see rightmost spot) shows the effective  measurement noise has abruptly changed, however the prior assumed by the HDP-SLDS segmentation only allowed the components of $\RR$ to change modestly.  
    The default settings of the MATLAB code associated with Ref. \cite{Fox2011} assumes little dispersion about the nominally known mean $\RR$ (all segments yielded an effective measurement noise having  $\approx 40.0nm$ on all diagonal components via the HDP-SLDS method).  
      Using a so-called ``non-informative'' or ``uniform'' prior can potentially remedy the situation, but such a prior can introduce 
      both technical and computational problems in nonparametric Bayesian methods \cite{GhoshNonparBook};  see S5 Text for further discussions on this issue.

       Alternatively, if one uses the segmentation afforded by the HDP-SLDS, but instead computes the MLE using a variant of the technique of Ref. \cite{Calderon2013b} (discussed in S1 Text) as opposed to using the posterior sample provided by the HDP-SLDS (which contains biases induced by priors), the $p-$values  were found to be $0.31$ and $0.45$ for the left and right segments, respectively.  The primary
       difference in the estimation results was in the noise components estimated, $\RR$ and $D$, the effective measurement noise and diffusion coefficient matrices, respectively  (see S3 Text).  
       For example, the MLEs  $\hat{\RR}=\mathrm{diag}([19.5^2,39.9^2])$ [$nm^2$] \& $\hat{D}=\mathrm{diag}([1.2 \times 10 ^{-1}, 1.4
       \times 10 ^{-2}])$ [$\mu m^2/s$] were obtained in the left segment and 
         $\hat{\RR}=\mathrm{diag}([140.1^2,93.9^2])$ [$nm^2$] \& $\hat{D}=\mathrm{diag}([4.8 \times 10 ^{-5}, 4.5 \times 10 ^{-5} ])$ [$\mu m^2/s$] were obtained in the right segment (i.e., the segment occurring later in time)
        using the model in Ref. \cite{Calderon2013b}. Hats are used to denote MLE estimates and $\mathrm{diag}(\cdot)$ denotes the square diagonal matrix formed by the arguments.   Note that the diffusion coefficient estimate, $\hat{D}$, is near machine single-precision zero in the right segment.
      
       The GFP image stack associated with this video readily shows photobleaching occurs  (see rightmost  point spread function spot  in S1 Data). 
       The MLE predicted effectively zero diffusion for the  right  segment (i.e., the state appearing temporally after the first segment), quantitatively indicating that significant photobleaching had occurred and that the second state was mainly background photon noise not corresponding to an individual molecule.  
       The  quantitative evidence for this statement is the MLE parameter estimates (effective measurement noise near diffraction limit in the ``right segment'') and the corresponding $p-$values, reported in two paragraphs above, resulting from testing the fitted model against the data.  
       In this experiment, the HDP-SLDS approach was able to automatically detect this measurement noise transition despite a poorly tuned prior;  the state change was identified without requiring subject matter expertise or manual image inspection.

    Fig. \ref{fig:twostate} displays another experimental trajectory with a different type of state switching which was readily identified by the HDP-SLDS procedure.  The posterior mode computed using Gibbs sampling  of the HDP-SLDS model contained assignments with two and three states;  in total $10^4$ posterior MCMC samples were sampled. A majority of the MCMC draws implied  two-state switching.  However, one does not necessarily need to be concerned about the number of states if one is only interested in determine change points.  The change points predicted by the HDP-SLDS are denoted by vertical lines (change points  were determined to occur if a state change occurred at time $i$ in over 25\% of the posterior samples).  Accurately identifying change points can help in quantifying the transition times between distinct physical states \cite{Persson2013}.
    However the state switching in this particular system was likely caused by vibrations in the microscope's piezo stage. 
    The HDP-SLDS can be used to automatically identify physically relevant state changes or state changes induced by experimental artifacts.  As we show in   Fig. S5, isolated outliers (e.g., caused by intense fluorescent background ``flashes'' in some SPT  experiments) are also readily identified by the HDP-SLDS framework.

    Next we turn to an example exhibiting subtle state changes (identified using the procedure outlined in S1 Text) in a more biologically relevant study.  
     Fig. \ref{fig:Fhist1} shows two GFP tagged chromatid trajectories during the metaphase stage of mitosis. The white light image of these trajectories is shown in Fig. \ref{fig:mitosis} where the spatial location of the RFP tagged mitotic spindle pole bodies (SPB) as well as the  GFP tagged chromatid trajectories are shown.  HDP-SLDS state segmentation was carried out for each of the trajectories; the state segmentation is denoted by vertical lines in the bottom panel of plots in Fig. \ref{fig:Fhist1}.  The procedure used to infer forces  in Fig. \ref{fig:Fhist} was carried out again for the pair of chromatids undergoing mitosis.  The histogram of the force magnitudes is not atypical for trajectories observed, however the evolution of  the force vectors was suggestive of a dynamic regime shift.

    In order to probe this regime shift further, we computed  the eigen-values and eigen-vectors  of the MLE of $\BB$ (a matrix associated with the continuous-time overdamped Langevin model;   details provided in S3 Text).  
    The eigen-analysis shown in Fig. \ref{fig:mitosis} provides quantitative information about variations in the effective force vector as a function of position (i.e., the eigen-analysis provides both magnitude and directional information about a linear vector field).  The upper-right panel shows that the largest magnitude eigen-value consistently decreased over time for each state identified by the HDP-SLDS segmentation. The rate of decrease was similar for the pair of sister  chromatids. This phenomenon was originally identified using the crude time window segmentation of Ref. \cite{Calderon2013b}, but the HDP-SLDS provides a framework for more systematically dividing the trajectory into different dynamical segments.

    In terms of physical interpretations, several molecular components can induce ``force cross-talk'' between sister chromatids.  A potential explanation of  the eigen-value phenomenon described above is that 
     microtubules of the mitotic spindle connected to the common centromere of the sister chromatids are simultaneously inducing a common force change to both chromatids.  
     Alternatively (or perhaps in addition), a common tension signal induced by chromatin loops and a network of pericentric proteins (cohesin and condensin \cite{Stephens2013b,Stephens2013a}) might be influencing the forces experienced by the pair of sister chromatids; recall that the GFP dyes used are in close proximity to the centromere so molecular changes in the pericentric region can potentially influence the fluctuations observed in the tagged sister chromatids \cite{Stephens2013b,Stephens2013a}.

    Regardless of the underlying biological cause of the phenomenon observed, the downstream eigen-analysis  substantially benefited from the temporal segmentation provided by the HDP-SLDS framework.  
      Using the segmentation afforded by the HDP-SLDS method resulted in the eigen-vectors shown in the bottom panel of Fig. \ref{fig:mitosis};  in this figure, the eigen-vectors associated with the largest magnitude eigen-value is displayed as a solid line and the  eigen-vector associated with the smaller eigen-value of the estimated 2D 
    $\BB$ matrix is shown by a dashed line (the origin for the eigen-vectors correspond to the empirical average of the identified states). 
     The (real-valued) eigen-values and eigen-vectors provide spatial information about restoring force directions experienced by the tagged chromatid.
     Observe how the pair of chromatids both exhibit a state change occurring around 16$s$. Note that the time of the change point  occurring around 16$s$ coincides for the sister chromatids, but each trajectory was processed independently by separate HDP-SLDS analyses.  At 16$s$, the 
     dominant eigen-vectors of the chromatids change from pointing towards the spindles to a direction nearly orthogonal to the plane connecting the two mitotic spindle poles. 
     The eigen-values for both chromatids are decreasing at roughly the same rate (i.e., their effective ``stiffness" is increasing).  
     Also, recall that the pair of chromatids are subject to forces from multiple sources that can change in magnitude and direction;  for example, tugging induced by the microtubules connected to the centromere shared by the chromatids, forces associated with nucleosome remodeling (e.g., RNA polymerase ``tugging'' on the strand as it makes RNA), as well as other spring like forces induced by the shared centromere and percentric components \cite{Stephens2013b,Stephens2013a}. 

     The various explanations posited above are admittedly speculative.  What these results unambiguously show is that sister chromatids experience a force regime shift whereby the direction and magnitude of the forces (inferred by position vs.~time data) co-occur in both direction and magnitude.  This information gives new insight into  
     the effective forces experienced by particles in time changing, complex live cellular environment. Researchers can get a more reliable ``force map'' that contains rich dynamical information beyond that contained in a mean square displacement or correlation-based analysis.  Note that this ``force map'' allows a spatially dependent vector field to change over time (an important feature for detecting dynamical changes in single-molecule measurements).  
     A dominant force direction pointing orthogonal to the spindle axis is highly suggestive of a new dominant force (not induced by the microtubule pulling) being felt by both sister chromatids.
     Deciding the correct  physical explanation of the phenomenon observed  requires additional experiments, but the results obtained on the metaphase trajectories are presented in order to show the power of SPT analysis coupled with HDP-SLDS segmentation and downstream likelihood-based time series analysis \cite{Calderon2013b}.   
     Currently we are attempting to use this type of analysis to more systematically and quantitatively 
     probe forces associated with the various physical phenomena described here using experiments which more carefully control the types of forces that can be experienced by the GFP tagged chromatids.

      Before proceeding, we should note that the GFP array used to tag the chromatids induces a gradual continuous trend in the
      effective measurement noise (the finding can readily  be observed in S1-S3 Data).  A gradual time trend induced in large part by different fluorophores bleaching out has been observed and quantitatively studied in previous works (see Supp. Mat. of Ref. \cite{Calderon2013b}).  However, the HDP-SLDS formulation considered assumes a constant measurement noise for the entire duration of the trajectory.  S6 - S7 Figs. explore various simulations mimicking features in the experimental data;  specifically we analyze data where the effective measurement changes linearly from 15nm to 35nm over the course of 400 measurements (this induces model misspecification since the HDP-SLDS model assumes fixed covariance). The linear change in measurement noise is consistent with the MLE range of values observed in S3 Data.
          Figs. S6 - S7 indicate that if one uses a prior near the temporal midpoint of the time changing measurement noise that the HDP-SLDS segmentation can achieve accuracy similar to the ideal situation where the HDP priors match the DGP exactly.   For whatever system is being studied, it is highly recommended to employ simulations mimicking features implied by the data in order to be more confident that the segmentation output does not contain artifacts induced by features not explicitly accounted for in the underlying model.   Techniques accounting for gradual time trends in certain parameters extending the HDP-SLDS  model would be an interesting future research topic.

      In terms of more general applicability of this type of analysis, systematically segmenting SPT trajectories into distinct dynamical regimes and then using classical statistical physics models is one alternative to using so-called anomalous diffusion models commonly used in SPT analysis 
     \cite{klafter08,Hajjoul2013, Weber2012}.  This is advantageous since fitted overdamped Langevin model output can readily be interpreted in terms of force, velocity, and molecular friction.  Furthermore, an overdamped Langevin SDE model  can be tested directly against experimental data and  the effects of measurement noise can be systematically  accounted for by likelihood methods \cite{SPAdsDNA,SPAfilter,Calderon2013b,arxivDec2013}. Anomalous diffusion models are more difficult to physically interpret and rigorous time series analysis (accounting for measurement noise, sampling noise, spatially dependent forces, etc.) is challenging for various technical reasons \cite{Calderon2013b}.

\section{Discussion}
\label{sec:discussion}

We have demonstrated the utility of the HDP-SLDS method of Fox et al. \cite{Fox2011} in automatically segmenting SPT data into different dynamical regimes. The approach shows great promise in systematically processing a variety of SPT data sets. 
When applied to experimental data, we demonstrated how new quantitative information can be extracted about  transient forces experienced during mitosis in live yeast cells; the method  explicitly accounts for the statistical effects of measurement noise on top of ``thermal'' or ``process'' noise (both sources are non-negligible in many SPT applications).  
% In  cases where photobleaching abruptly altered noise statistics at an unknown change point, we demonstrated the utility of the HDP-SLDS method  and discussed advantages of using the approach.
     Simulations results reported in S2 Text were used to exhibit strengths and weaknesses of the HDP-SLDS approach within an SPT context.  Work presented in Ref. \cite{Calderon2014} compared the HDP-SLDS approach to  classic (finite state space) HMM models.

    In terms of strengths, the HDP-SLDS  approach can account for an unknown number of states, position dependent anisotropic active forces, and can account for the statistical effects of  measurement noise by exploiting Kalman filtering modeling in conjunction with HDP ideas \cite{Teh2006a};  these features are not accounted for in any HMM approach currently used in SPT analysis and hence the HDP-SLDS represents a ``state-of-the-art'' technology within this domain.  The approach provides a more systematic means for the ``time window size'' selection problem discussed in Ref. \cite{Calderon2013b}.  The HDP-SLDS approach was also confirmed to be fairly robust to many modeling assumptions and hyperparameters associated with HDP modeling \cite{Fox2011}.  
    In situations where a collection of transient dynamical responses are experienced   within a single trajectory and the ``states'' causing the different kinetics are 
    experimentally resolvable,
    the data processing procedure discussed here  provides an attractive alternative to anomalous diffusion modeling
     \cite{klafter08,Hajjoul2013,Calderon2013b, Calderon2013,Weber2012} since
    the procedure 
      can mitigate artifacts  induced  by aggregating  distinct kinetic states \cite{Weber2012,Calderon2013b}.  Furthermore, the HDP-SLDS approach can be modified to
      produce output that can be readily physically interpreted in terms of classic SPT models (see Sec.\ref{sec:ODdesc}).

    A substantial weakness of the HDP-SLDS approach (also discussed in Ref. \cite{Fox2010}) is associated with its dependence on  reliable priors for the parameters determining governing the Kalman filter parameters (i.e., hyperparameters specifying the ``base measure'' \cite{Fox2010}).  The data-driven scheme advocated in Ref. \cite{Fox2011} for selecting the prior mean of $\RR$ exhibits undesirable properties in many SPT applications (results presented in S2 Text).  
    A variant of the MLE-based approach of Ref. \cite{SPAfilter,Calderon2013b} was discussed (S1 Text) and was demonstrated (S2 Text)  to be capable of providing pilot estimates needed to tune the base measure hyperparameters priors in SPT applications.  
     The formal hypothesis testing methods utilized in Ref. \cite{SPAfilter,Calderon2013b} also allow one to test the validity of a candidate segmentation and assumed dynamical model against experimental data in situations where reference (or ``ground-truth'') data is unavailable.

    In summary, combining recent nonparametric Bayesian modeling ideas, like the HDP-SLDS, with frequentist ideas  show great potential in quantitatively analyzing complex SPT data sets. In the future, we aim to provide user-friendly software making the tools discussed herein readily accessible to the diverse set of researchers involved in SPT.  The development of new, statistically rigorous data processing algorithms capable of reliably extracting information about molecular motion from live cell data is becoming ever more important  as \emph{in vivo} microscopy techniques continue to improve in spatial and temporal resolution \cite{Thompson2010,Ram2012,Gao2014,Welsher2014,Chen2014}.

% Do NOT remove this, even if you are not including acknowledgments.

\section*{Acknowledgments}

CPC was supported by internal R\&D funds from Ursa Analytics and KB was funded by the National Institutes of Health R37 grant GM32238. 

\bibliographystyle{plos2009}

\newpage

%before submitting, commment out all figure stuff except \caption info
\section*{Figures}

% This section is for figure legends only, do not include
% graphics in your manuscript file.
%
\renewcommand{\figurename}{Fig} %adhere to PLOS's new 2015 format
\setcounter{figure}{0}
\makeatletter 
\renewcommand{\thefigure}{\@arabic\c@figure}
\makeatother

\begin{figure}[H]
\center
\centering
\begin{overpic}[width=0.95\textwidth]{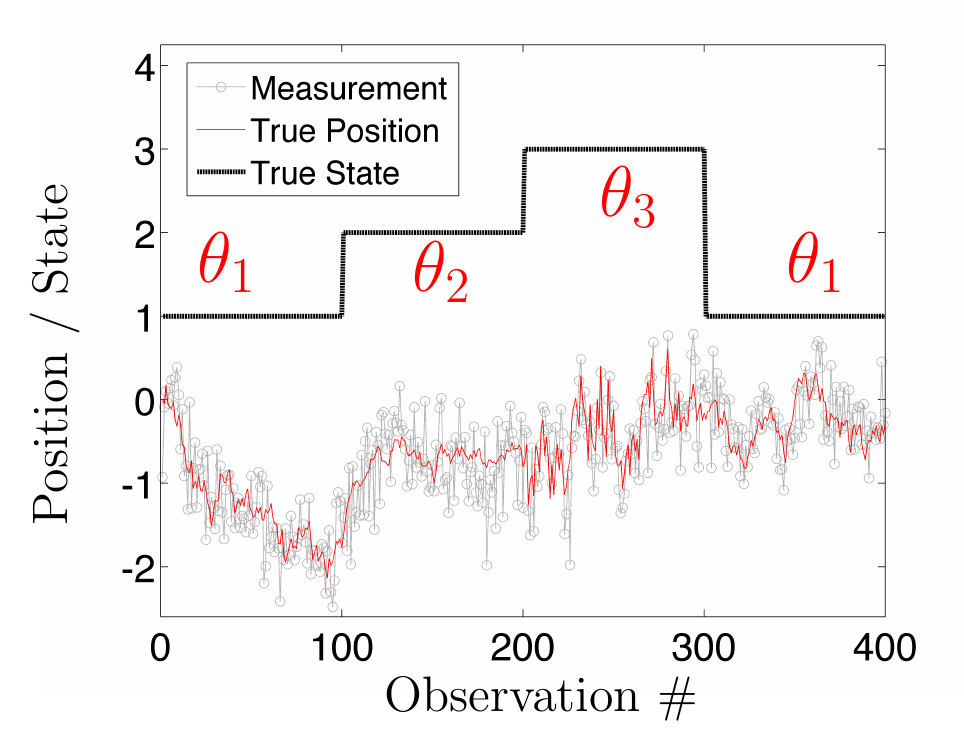}
\end{overpic}
%note: combination of "\textbf{.}" with caption package labelsep=none is messy but only way i could get inline \refs and 
%mandated formatting to jive.
\caption{\textbf{.}
\footnotesize  
Illustrative example of a Switching Linear Dynamical System (SLDS). The stochastic dynamics of the (simulated) particle is denoted by thin red line (i.e., the trajectory of $\vec{r}$) and the simulated discrete measurements, $\vec{\psi}$,  
are represented by the grey lines with symbols.  The parameter determining the linear dynamical system is denoted by $\theta_z$.  At random times, the latent state $z$ changes and hence the underlying parameter specifying the dynamics, $\theta_z$,  also changes.  
Note that $\theta_z$ models both thermal and measurement noise (parameters defined in Eqn. \ref{eq:SDE}).  The latter is crucial for identifying subtle changes in the underlying stochastic process.  For example, the dramatic change in thermal fluctuations shown around observations 300-400 would be difficult to detect without jointly modeling thermal and measurement noise. To achieve this joint modeling, one must also account for temporal correlations induced by measurement noise \cite{SPAfilter,Berglund2010,Calderon2013}.      
}
\label{fig:illEG}
\end{figure}

\clearpage
\newpage

\begin{figure}[htb]
\center
\centering
\begin{overpic}[width=0.95\textwidth]{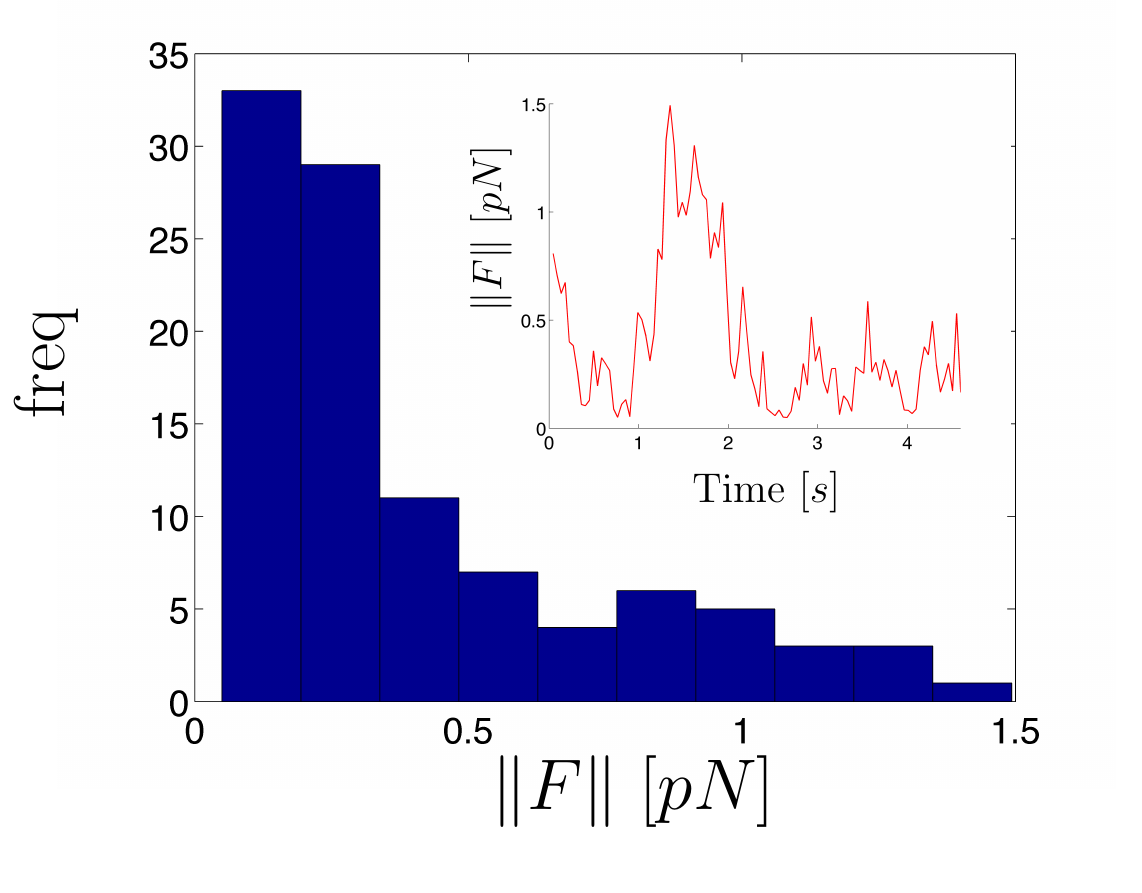}
\end{overpic}
\caption{ \textbf{.} 
\footnotesize 
 Histogram of inferred force magnitudes  computed from experimental chromatid position vs.~time data. The inset displays the  force trajectory.  Forces were computed using HDP-SLDS state segmentation and Kalman filtering (see text for details). The white light image of cell with the position vs.~time data measured from the GFP channel is shown in  Fig. \ref{fig:noiseJump}.
}
\label{fig:Fhist}
\end{figure}

\clearpage
\newpage

\begin{figure}[H]
\center
\centering
\begin{overpic}[width=0.95\textwidth]{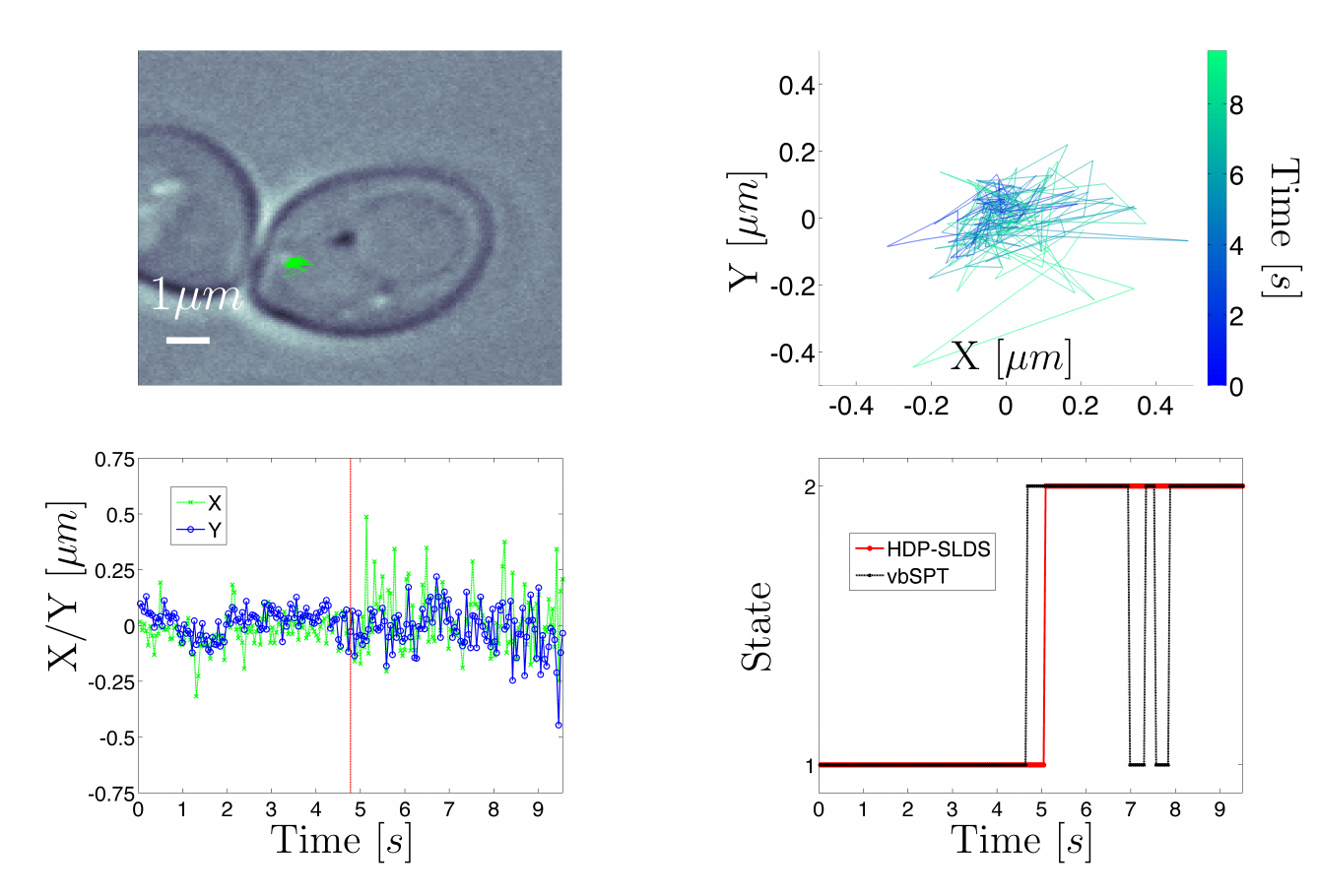}
\end{overpic}
\caption{ \textbf{.}
\footnotesize 
Experimental trajectory of GFP tagged chromatid in yeast.  Top left panel: White light image where  green lines denote the SPT trajectory measured via the GFP channel.  Top right panel: Zoomed snapshot of trajectory shown in white light image;  time is color coded.  Bottom left panel:  $X/Y$ vs. time for trajectory.  Vertical red line denotes the time point at which the HDP-SLDS inferred a state change.  Bottom right panel: HDP-SLDS \cite{Fox2011} and vbSPT \cite{Persson2013} state estimates. 
}
\label{fig:noiseJump}
\end{figure}

\clearpage
\newpage

\begin{figure}[H]
\center
\centering
\begin{overpic}[width=0.95\textwidth]{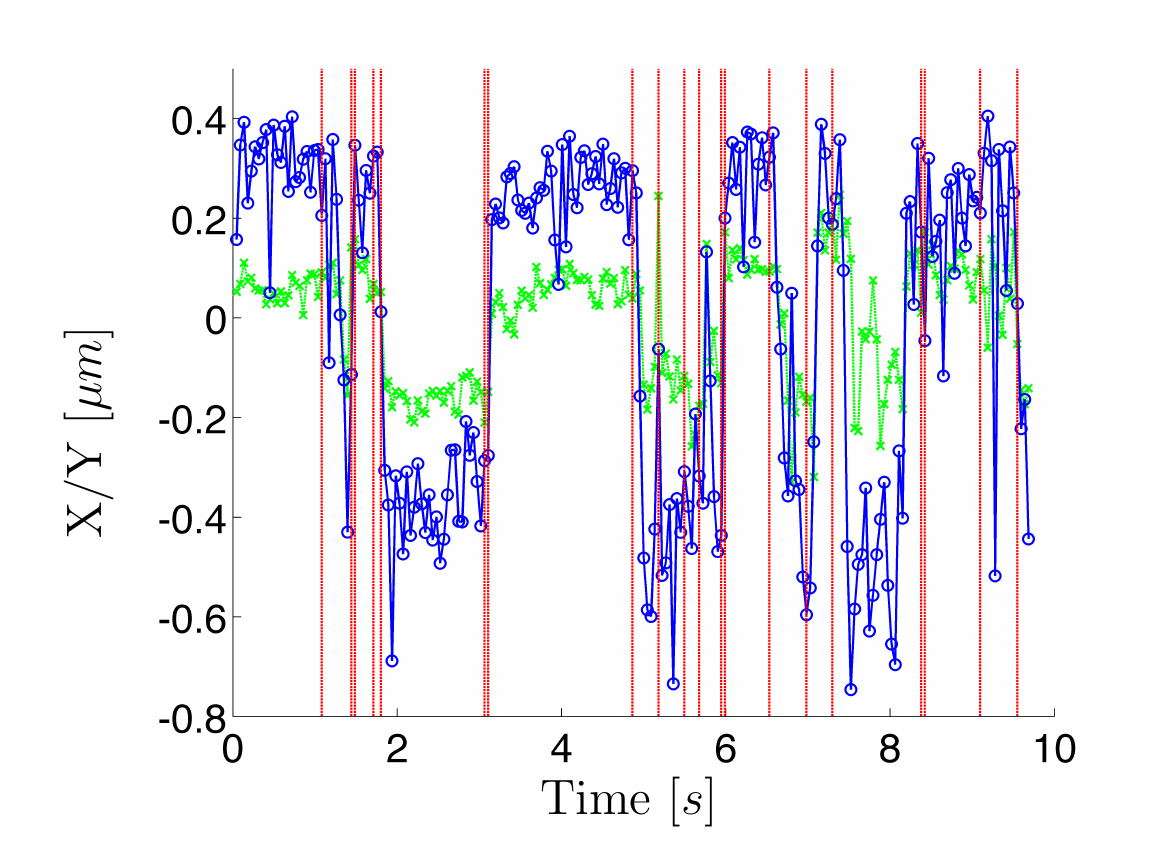}
\end{overpic}
\caption{ \textbf{.}
\footnotesize 
A GFP tagged 2D chromatid trajectory exhibiting rapid two state regime switching.  The vertical red lines denote the change points identified by the HDP-SLDS analysis.  
}
\label{fig:twostate}
\end{figure}

\clearpage
\newpage

\begin{figure}[htb]
\center
\centering
\begin{overpic}[width=0.95\textwidth]{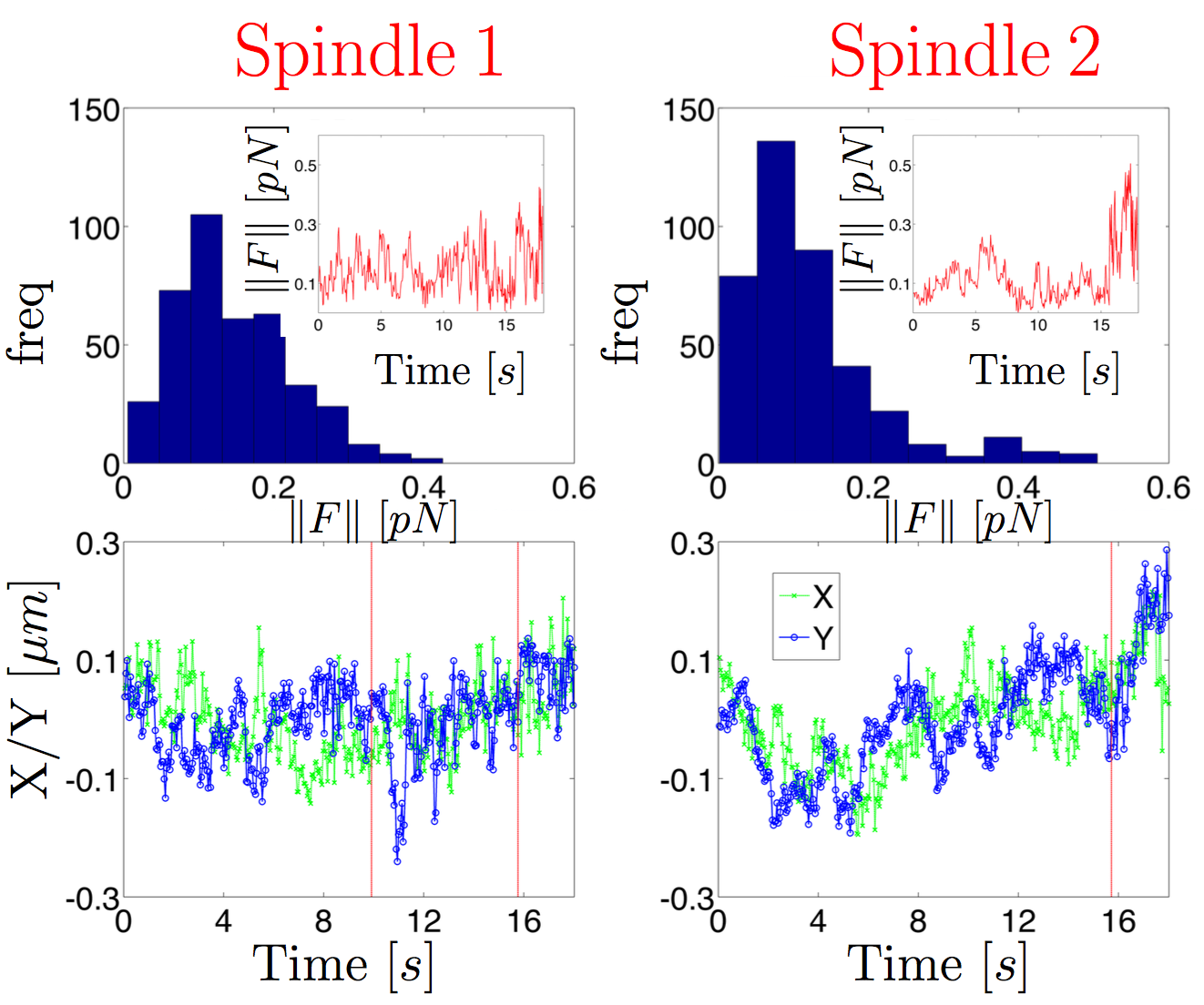}
\end{overpic}
\caption{ \textbf{.}
\footnotesize 
Analysis of two sister chromatid trajectories in the metaphase of mitosis (each column shows results of one chromatid).   Spatial location of the spindle pole bodies (SPB) and the two sister chromatid trajectories are displayed in Fig. \ref{fig:mitosis}.  The top row displays the forces inferred using the Kalman filter analysis discussed in the main text.  The bottom row displays the measured trajectories vs. time as well as the time point of the state changes predicted by the HDP-SLDS approach.   
}
\label{fig:Fhist1}
\end{figure}

\clearpage
\newpage

\begin{figure}[H]
\center
\centering
\begin{overpic}[width=0.95\textwidth]{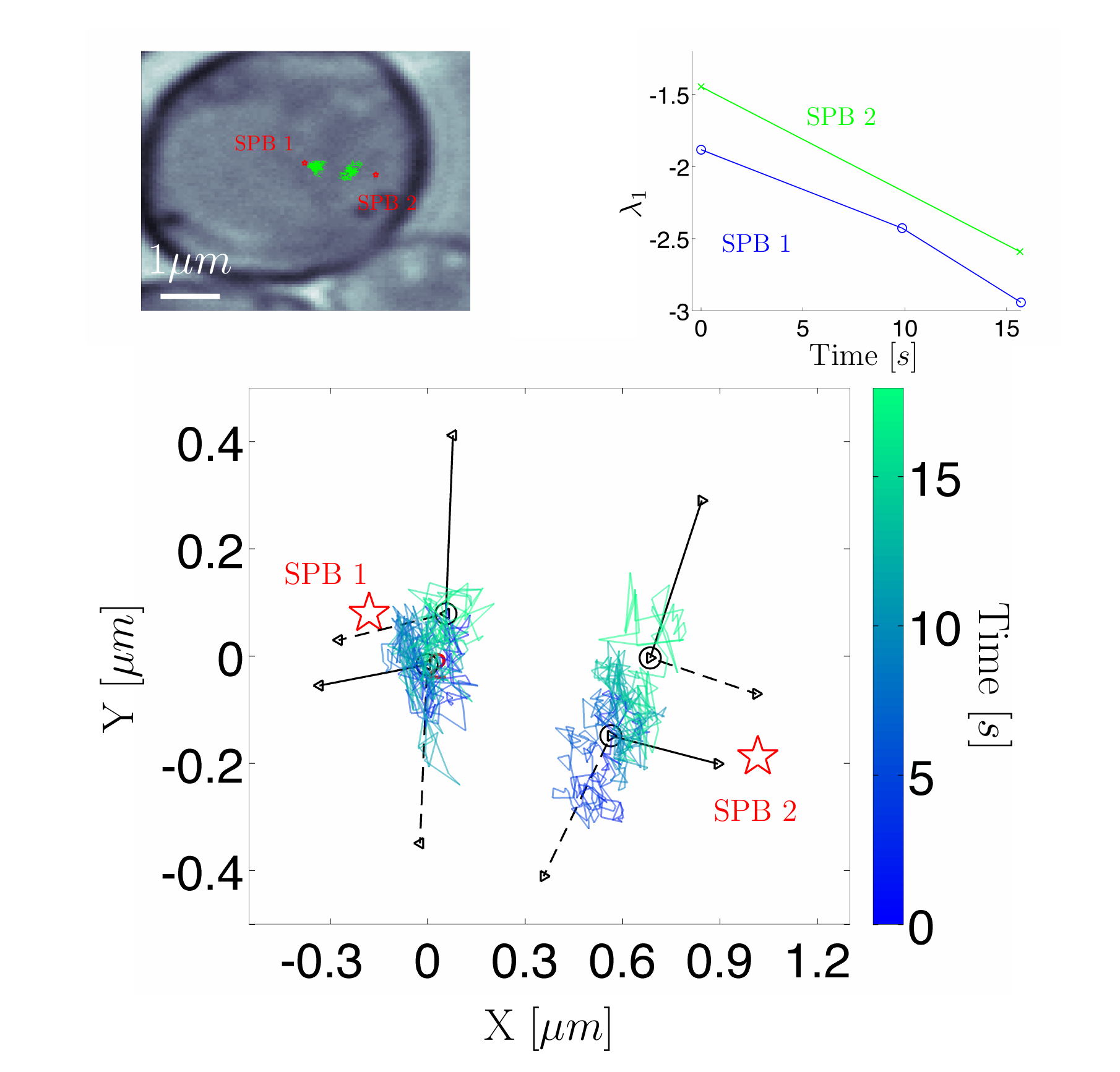}
\end{overpic}
\caption{ \textbf{.}
\footnotesize 
Top left panel:  white light image of two chromatid trajectories in yeast;  spindle pole body (SPB) locations denoted by red.  Top right panel: the largest eigen-value (quantifies effective restoring force experienced by chromatid) vs. time;  a new restoring force was inferred for each of the unique states identified by the HDP-SLDS methods (three unique states for SPB 1 and two states for SPB 2).
Bottom panel: Eigen-vectors implied for the various states;  solid line denotes the eigen-vector associated with the largest eigen-value and the dashed line denotes the weaker eigen-vector.  For spindle 1, we display the eigen-vectors associated with state 2 and 3 (state 1's eigen-vectors were similar to state 2, so these vectors were omitted for clarity).
}
\label{fig:mitosis}
\end{figure}

\end{document}